\documentclass[twocolumn,brazil,english,pra,aps,amssymb,amsfonts,floatfix]{revtex4}
\usepackage[T1]{fontenc}
\usepackage[latin1]{inputenc}
\usepackage{amsmath}
\usepackage{graphicx}
\usepackage{amssymb}

\makeatletter
\usepackage{graphicx}
\usepackage{amscd}
\usepackage{bm}

\usepackage{babel}
\makeatother
\begin{document}

\title{Dissipative Stern-Gerlach recombination experiment}

\author{Thiago R. Oliveira and A. O. Caldeira}

\affiliation{Departamento de F\'{\i}sica da Mat\'{e}ria Condensada, Instituto
de F\'{\i}sica Gleb Wataghin,Universidade Estadual de Campinas}

\affiliation{Caixa Postal 6165, Campinas, SP, CEP 13083-970, Brazil}

\begin{abstract}
The possibility of obtaining the initial pure state in a usual Stern-Gerlach
experiment through the recombination of the two emerging beams is
investigated. We have extended the previous work of Englert, Schwinger
and Scully \cite{ISG1} including the fluctuations of the magnetic
field generated by a properly chosen magnet. As a result we obtained
an attenuation factor to the possible revival of coherence when the
beams are perfectly recombined . When the source of the magnetic field
is a SQUID(superconducting quantum interference device) the attenuation
factor can be controlled by external circuits and the spin decoherence
directly measured. For the proposed SQUID with dimensions in the scale
of microns the attenuation factor has been shown unimportant when
compared with the interaction time of the spin with the magnet.
\end{abstract}

\email{tro@ifi.unicamp.br}

\maketitle

\section{Introduction}

The Stern-Gerlach experiment gives an experimental evidence of the
quantum nature of the spin of a particle. When a beam of spin 1/2
particles, in the eigenstate $|+\rangle$ of $S_{x}$, goes through
a variable magnetic field in the $\hat{z}$ direction and the outcoming
particles are detected on a screen, we observe the presence of two
distinct peaks corresponding to the spins in the positive and negative
$\hat{z}$ direction.

Moreover, besides giving an evidence of the spin quantization, this
example is considered the paradigm of the measurement process. The
beam goes into the magnet in a pure state, for example, the eigenstate
$|+\rangle_{x}$ of $S_{x}$, and as far as the measurement of $S_{z}$
is concerned it is described as a superposition of the eigenstates
of $S_{z}$. Later, as a result of the measurement process, the quantum
state collapses into one of the eigenstates of the latter.

A question that arises is: could we get the initial pure state of
the system if we recombine the two beams from the Stern-Gerlach experiment?

This idea of the recombination of the beams, through the so-called
Stern-Gerlach Interferometer(SGI), is an old one, but has always been
treated in a qualitative way \cite{D.Bohm}. One of the first quantitative
studies was presented in a series of three articles \cite{ISG1,ISG2,ISG3},
in which the authors examined the possibility of recombination and
concluded that the precision of control of the magnetic field is of
fundamental importance to the reconstruction of the initial pure state.
However, in those articles the authors did not take into account the
origin of the magnetic field and its possible sources of fluctuations.

In a more recent study \cite{Rodrigo} this problem has been revisited
and the quantum nature of the magnetic field taken into account. However,
only the coupling of the uniform part of field to the spin of the
particle has been considered and the effects of its fluctuations on
the spatial part of the wave function have been disregarded.

In another manuscript \cite{Banerjee} a model for a dissipative Stern-Gerlach
experiment was proposed without explicit reference to the origin of
the fluctuations. It basically dealt with a Stern-Gerlach apparatus
inside a dissipative medium and, besides, did not consider the recombination
process.

In this manuscript we intend to make a fully quantum description of
the magnetic field and study the effects of its unavoidable fluctuations
in the spatial part of the wave function. We also pay special attention
to the origin of these fluctuations by presenting a model where the
magnetic field is generated by a pair of SQUIDs where the magnetic
flux is known to obey a Langevin equation. We will be only interested
in the spin coherence that can be measured through the mean value
of $S_{x}$ and arises from the off-diagonal terms of the reduced
density operator of the system in the spin space.

Another point that is worth mentioning here is the fact that we will
be aiming at a problem quite different from the standard one in the
area of dissipative systems. Usually people want to get rid of decoherence
for macroscopic quantum mechanical variables whereas we will try to
increase the effect of decoherence on a microscopic variable. The
reason for this is to fully test the existing theory of decoherence
as unequivocally due to the sources of noise and/or dissipation known
to be coupled to the system.

The paper is organized as follows: in Sec. $\mathrm{II}$ we present
a model for the SGI where the magnetic field is generated by a pair
of SQUIDs. In Sec. $\mathrm{III}$ we evaluate the dynamics of the
reduced density operator in the spin space using the Feynman-Vernon
path integral approach. In Sec. $\mathrm{IV}$ we analyze the coherence
revival in the SGI. Finally, we summarize our results in Sec $\mathrm{V}$.

\section{The Stern-Gerlach Interferometer}

The SGI can be divided in two parts: the first one is responsible
for separating the beams and the second one for recombining them.
We will follow the basic model and approximations of \cite{ISG1},
considering that the beams split in the $\hat{z}$ direction due to
the magnetic field gradient which also points along this direction.
As referred to a given origin, half way from two specific magnets
(see arrangement below), the magnetic field should depend on \textit{y}
in an anti-symmetric way to enable a perfect recombination of the
two beams. In order to solve the problem we will consider that the
velocity in the $\hat{y}$ direction is constant which allows us to
replace its position dependence by a time dependence.

Our model Hamiltonian is\begin{equation}
H=\frac{p^{2}}{2m}-\sigma_{z}f\left(t\right)z\,,\end{equation}
 where $f\left(t\right)=\mu\,\partial B/\partial z$ and $\mu$ is
the magnetic moment of the particle. Here we are not considering the
uniform part of the magnetic field that is only responsible for the
precession the spin about the $\hat{z}$ direction and has already
been treated in \cite{Rodrigo}.

For a perfectly noiseless anti-symmetric field with a piecewise constant
gradient we observe loss and revival of the spin coherence through
the behavior of the off-diagonal elements of the reduced density operator
in the spin space, as shown in Figure 1. There, $T$ is the experiment
time and the revival in the middle of the experiment is due to the
recombination of the wave packets in the momentum space. The vertical
lines show four different regions of a SGI. In the first part the
beams are accelerated in opposite directions along $\hat{z}$. The
second and third parts are due to the empty interval between the magnets
where the direction of the acceleration for each beam suddenly changes.
In the last part the acceleration of each beam returns to its original
value and they are finally recombined as they emerge from the set
up.

\begin{figure}
\includegraphics[%
  scale=0.8]{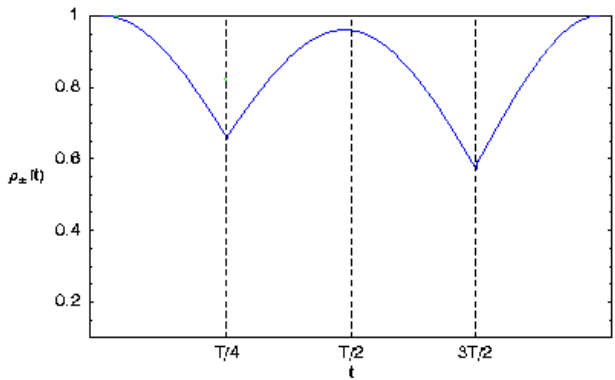}

\caption{Loss and revival of coherence in a SGI without fluctuations.}
\end{figure}

We now introduce the fluctuations of the magnetic field to the problem.
In order to do this we will consider a SGI where each of the two magnets
is formed by a pair SQUIDs arranged as shown in figure 2 . With this
model we obtain a field configuration very similar to that proposed
in \cite{ISG1} and described above.

\begin{widetext}

\begin{figure}[h]
\includegraphics[%
  scale=0.5]{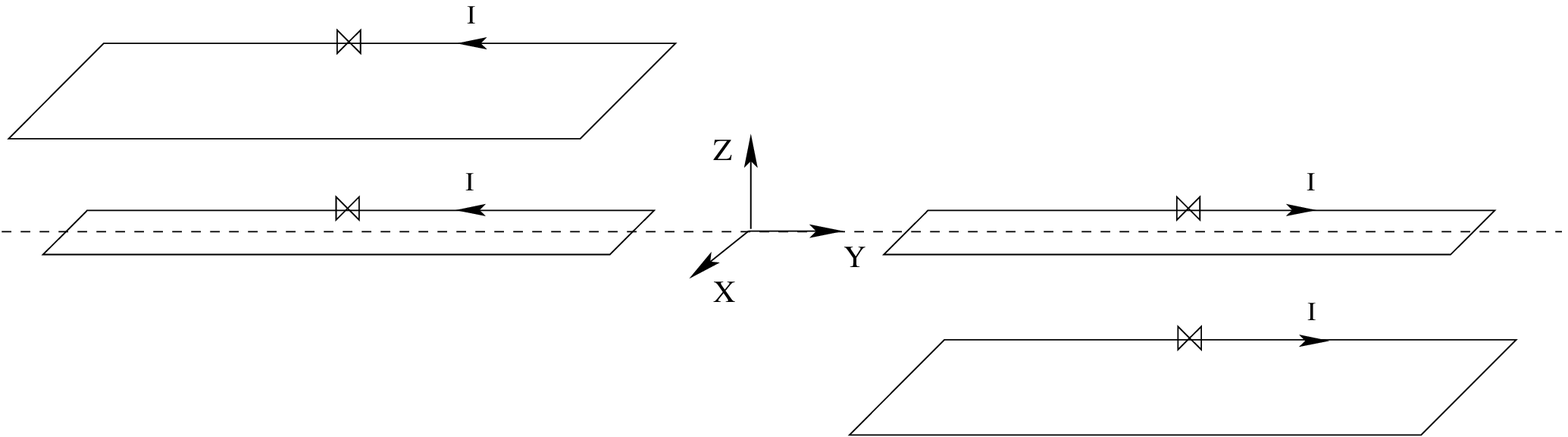}~~~~~~~~\includegraphics[%
  scale=0.75]{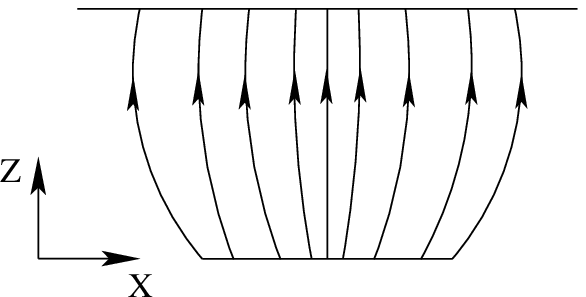}

\caption{Lateral view of the proposed apparatus for the ISG and the magnetic
field lines in a frontal view.}
\end{figure}
\end{widetext}

The total flux inside a SQUID has its dynamics described by the model
Hamiltonian given by \cite{Caldeira}

\begin{eqnarray}
H_{sq+osc} & = & \frac{P_{\phi}^{2}}{2C}+\frac{\Phi'^{2}}{2L_{0}}+\sum_{k}\left[\frac{P_{k}²}{2m_{k}}\right.\nonumber \\
 &  & \left.+\frac{m_{k}\omega_{k}^{2}}{2}\left(x_{k}+\frac{C_{k}}{m_{k}\omega_{k}^{2}}\Phi'\right)^{2}\right]\,,\label{eq 2}\end{eqnarray}
 with the spectral function $\mathcal{J}\left(\omega\right)=\frac{\pi}{2}\sum_{k}\frac{C_{k}^{2}}{m_{k}\omega_{w}}\delta\left(\omega-\omega_{k}\right)=\eta\omega$.
Here $\Phi'$ is the fluctuation about one of the many metastable
flux values at zero external field. We are considering that there
is no tunnelling or thermal activation of the flux variable to its
neighboring minima. Thus, the flux is oscillating around a local minimum
close to $n\Phi_{0}$ and $L_{0}=\left[1/L+2\pi i_{0}/n\Phi_{0}\right]^{-1}$
is an effective inductance, that arises from a second order expansion
around the minimum of the electromagnetic potential energy. $i_{0}$
is the critical current for the SQUID. For this approximations to
be valid we should be sure that we have many minima, in other words:
$2\pi i_{0}L/\Phi_{0}\gg1$ . As a matter of fact we shall assume
that each SQUID is carrying a persistent current corresponding to
the metastable minimum at $n=1$.

Here, it should be emphasized that we are only taking into account
the parameters of one single SQUID of each pair, say, the smaller
one. Operationally it means that the particle trajectories are always
close to that same circuit over half of the experiment time. In the
second half of the experiment the same applies to the second pair
of SQUIDs. The presence of the larger SQUID in each pair is only to
mimic the field lines of a usual Stern-Gerlach apparatus. We believe
that this requirement can be dropped at the expense of using effective
parameters for the coupled SQUIDs in a Hamiltonian of the same form
as (\ref{eq 2}).

Therefore, we are dealing with a particle coupled to a magnetic field
that is produced by a flux which, on its turn, is coupled to a bath
of harmonic oscillators. Now if we write the magnetic field in terms
of the flux we will have the particle indirectly coupled to the bath
through the flux. This can be done if we write $\Phi=B\left(z\right)A\left(z\right)$
where $B\left(z\right)$ is the magnetic field at $z$ as seen by
the particles whereas $A\left(z\right)$ is an effective area crossed
by the field lines through which the magnetic flux is exactly given
by the total value of $\Phi$ inside the ring. Since the latter does
not depend on $z$ we have

\begin{equation}
B\left(z\right)\frac{\partial A}{\partial z}+A\left(z\right)\frac{\partial B}{\partial z}=0\,,\label{eq:3}\end{equation}
 that can be manipulated to give

\begin{equation}
\frac{\partial B}{\partial z}=a\left(z\right)\Phi\,,\,\,\,\,\,\,\textrm{with}\,\,\,\,\,\, a\left(z\right)=\frac{1}{A\left(z\right)^{2}}\frac{\partial A}{\partial z}\,.\label{eq:4}\end{equation}

Using of the last expression, the total Hamiltonian reads\begin{eqnarray}
H_{part+sq+osc} & = & \frac{P^{2}}{2m}+\epsilon\sigma_{z}n\Phi_{0}z+\epsilon\sigma_{z}\Phi'z+\frac{P_{\phi}^{2}}{2C}+\frac{\Phi'^{2}}{2L_{0}}\nonumber \\
 &  & +\sum_{k}\left[\frac{P_{k}²}{2m_{k}}+\frac{m_{k}\omega_{k}^{2}}{2}\left(x_{k}+\frac{C_{k}}{m_{k}\omega_{k}^{2}}\Phi'\right)^{2}\right]\,,\nonumber \\
\label{eq:5}\end{eqnarray}
 with $\epsilon=\mu a\,$.

Now we have the particle coupled indirectly to the bath of oscillators.
Following the prescription of \cite{Ambegaokar} we can eliminate
the flux variable in the Hamiltonian and write a new one where the
particle is coupled directly to new oscillators with an effective
spectral function. The new Hamiltonian is

\begin{equation}
\tilde{H}=\frac{P^{2}}{2m}+\sigma_{z}f_{0}z+\sum_{k}\left[\frac{\tilde{P}_{k}²}{2\tilde{m}_{k}}+\frac{\tilde{m}_{k}\tilde{\omega}_{k}^{2}}{2}\left(\tilde{x}_{k}+\frac{\tilde{C_{k}}}{\tilde{m}_{k}\tilde{\omega}_{k}^{2}}\sigma_{z}z\right)^{2}\right]\end{equation}

For an original Ohmic spectral function, which is known to hold for
SQUIDs, we obtain the following effective spectral function:\begin{equation}
\mathcal{J}_{eff}\left(\omega\right)=\frac{\eta\omega}{1+\left(\frac{\omega}{\Omega}\right)^{2}+\left(\frac{\omega}{\Omega'}\right)^{4}}\end{equation}
 with $\eta=\epsilon^{2}L_{0}^{2}/R$, $\Omega=1/\sqrt{L_{0}^{2}/R^{2}-2CL_{0}}$
and $\Omega'=1/\sqrt{CL_{0}}$. Here $R$ and $C$ are, respectively,
the resistance and capacitance of the junction of the SQUID. The effective
spectral function is Ohmic with a new cutoff frequency given by the
minimum of $\Omega$ and $\Omega'$.

For typical SQUIDs we have the following values: $C\sim10^{-12}F,\, L\sim10^{-10}H,\, i_{0}\sim10^{-5}A,$
and $R\sim1\Omega$. To obtain $a\left(z\right)$ we will suppose
a SQUID with dimensions of $\left(10^{-5}\times10^{-3}\right)m^{2}$
and will estimate the total flux using the field in the middle of
two infinite wires which is multiplied by the area of $10^{-8}\, m^{2}$.
Comparing this with the flux at a distance $z$ we can obtain an expression
for $A\left(z\right)$ and then estimate $a\left(z\right)$. For $z=10^{-3}m$
we have $a\left(z\right)=10^{13}m^{-3}$. These specifications for
the apparatus give $L_{0}\sim10^{-10}H$ and $\eta=10^{-40}Kg\, s^{-1}$.
For cooper atoms, $m=1,8\times10^{-25}kg$, the relaxation time, $\gamma^{-1}$,
is of the order of $10^{15}$ seconds. 

We end this section with this proposal for a model for the SGI that
incorporates unavoidable sources of fluctuations of the magnetic field.
In our model these fluctuations are modelled by a coupling to a bath
of oscillators following \cite{Caldeira}. Now we have to evaluate
the reduced density operator of the system in the spin space to see
the behavior of the spin coherence through its off-diagonal elements.

\section{The reduced density operator}

Our model of a pair of SQUIDs generating the magnetic field for the
SGI obeys the following Hamiltonian\begin{equation}
H=H_{0}+H_{I}+H_{R},\label{eq:8}\end{equation}
 with \begin{eqnarray}
H_{0} & = & \frac{p^{2}}{2m}+\sigma_{z}f_{0}\left(t\right)z\,,\\
H_{I} & = & \sigma_{z}z\sum_{k}C_{k}x_{x}\,\,\,\,\,\,\,\textrm{and}\\
H_{R} & = & \sum_{k}\left[\frac{p_{k}^{2}}{2m_{k}}+\frac{m_{k}\omega_{k}^{2}}{2}x_{k}^{2}+\frac{C_{k}^{2}}{2m_{k}\omega_{k}^{2}}\left(\sigma_{z}z\right)^{2}\right].\end{eqnarray}

It describes the dynamics of a particle submitted to a linear potential
and also coupled to a bath of oscillators both in a spin dependent
way. The problem of a particle in a linear potential and coupled to
a bath of oscillators has already been treated in the literature and
the only difference here is the spin dependence that was absent in
\cite{Banerjee}. However, since we have only one spin operator in
(\ref{eq:8}) it acts as a parameter in our problem and we will use
a slightly modified Feynman-Vernon formalism to solve it. We are not
going to show all the details, of the calculation for which the reader
should consult \cite{Caldeira,Weiss,Thiago}.

For separable initial conditions, where the interaction between the
system and the bath is turned on at $t=0$, one has $\left\langle x'\mathbf{R}'S\right|\rho\left(0\right)\left|y'\mathbf{Q}'S'\right\rangle =\left\langle x'S\right|\rho_{1}\left(0\right)\left|y'S'\right\rangle \left\langle \mathbf{R}'\right|\rho_{2}\left(0\right)\left|\mathbf{Q}'\right\rangle $
and the four elements of the reduced density operator of the system
in the spin space are written as

\begin{equation}
\rho_{ss'}\left(x,y,t\right)=\int\int dx'dy'J_{ss'}\left(x,y,t;x',y',0\right)\left\langle x'S\right|\rho_{1}\left(0\right)\left|y'S'\right\rangle ,\end{equation}
 with \begin{widetext}

\begin{equation}
J_{ss'}\left(x,y,t;x',y',0\right)=\int\int\int d\mathbf{R}d\mathbf{R}'d\mathbf{Q}'K_{s}
\left(x,\mathbf{R},t;x',\mathbf{R}',0\right)\left\langle \mathbf{R}'\right|\rho_{2}\left(0\right)
\left|\mathbf{Q}'\right\rangle K_{s'}^{*}\left(y,\mathbf{R},S',t;y',\mathbf{Q}',0\right)\end{equation}

\end{widetext} which propagates the reduced density operator in time. In this expressions
$\mathbf{R},\mathbf{R}',\mathbf{Q}'$ are arbitrary configurations
(N-dimensional vectors) of the bath of oscillators. Within the path
integral formalism it is written as\begin{widetext}

\begin{equation}
J_{ss'}\left(x,y,t;x',y',0\right)=\int_{x'}^{x}Dx\left(t'\right)\int_{y'}^{y}
Dy\left(t'\right)e^{\frac{i}{\hbar}\left(S_{0}^{s}\left[x\left(t'\right)\right]-S_{0}^{s'}
\left[y\left(t'\right)\right]\right)}F_{ss'}\left[x\left(t'\right),y\left(t'\right)\right]
\end{equation}
with

\begin{eqnarray}
F_{ss'}\left[x\left(t'\right),y\left(t'\right)\right] & = & \int\int\int d\mathbf{R}d\mathbf{R}'
d\mathbf{Q}'\rho_{2}\left(\mathbf{R}',\mathbf{Q}',0\right)\int_{\mathbf{R}'}^{\mathbf{R}}
D\mathbf{R}\left(t'\right)\int_{\mathbf{Q}'}^{\mathbf{Q}}D\mathbf{Q}\left(t'\right)\nonumber \\
 &  & \times e^{\frac{1}{\hbar}\left(S_{I}^{s}\left[x\left(t'\right),\mathbf{R}\left(t'\right)\right]-
 S_{I}^{s'}\left[y\left(t'\right),\mathbf{Q}\left(t'\right)\right]\right)}e^{\frac{1}{\hbar}
\left(S_{R}^{s}\left[\mathbf{R}\left(t'\right)\right]-S_{R}^{s'}\left[\mathbf{Q}
 \left(t'\right)\right]\right)}\end{eqnarray}

\end{widetext} being the so-called influence functional, which contains all the
influence of the bath on the system. It has been evaluated before
\cite{Weiss} and the resulting expression is \begin{equation}
F_{ss'}\left[x\left(t'\right),y\left(t'\right)\right]=e^{-\frac{1}{\hbar}\Phi_{ss'}\left[x\left(t'\right),y\left(t'\right)\right]}\,,\end{equation}
 in which

\begin{widetext}\begin{eqnarray}
\Phi_{ss'}\left[x\left(t'\right),y\left(t'\right)\right] & = & \frac{im}{2}\left[Sx\left(0\right)-S'y\left(0\right)\right]\int_{0}^{t}dt'\,\gamma\left(t'\right)\left[Sx\left(t'\right)-S'y\left(t'\right)\right]+\nonumber \\
 & + & \frac{im}{2}\int_{0}^{t}dt'\int_{0}^{t'}dt''\,\left[Sx\left(t'\right)-S'y\left(t'\right)\right]\gamma\left(t'-t''\right)\left[S\dot{x}\left(t''\right)+S'\dot{y}\left(t''\right)\right]+\nonumber \\
 & + & \int_{0}^{t}dt'\int_{0}^{t'}dt''\,\left[Sx\left(t'\right)-S'y\left(t'\right)\right]\alpha_{R}\left(t'-t''\right)\left[Sx\left(t''\right)-S'y\left(t''\right)\right]\,,\label{eq:17}\end{eqnarray}

\end{widetext} with

\begin{eqnarray}
\gamma\left(t\right) & = & \frac{2}{m\pi}\theta\left(t\right)\int_{0}^{\infty}\frac{\mathcal{J}\left(\omega\right)}{\omega}\cos\omega t\,\, d\omega\end{eqnarray}
 and

\begin{equation}
\alpha_{R}\left(t'-t''\right)=\frac{1}{\pi}\int_{0}^{\infty}\mathcal{J}\left(\omega\right)\coth\left(\frac{\hbar\omega}{2KT}\right)\cos\omega\left(t'-t''\right)d\omega\,.\end{equation}

All we have to know is the spectral function of the bath,\[
\mathcal{J}\left(\omega\right)=\frac{\pi}{2}\sum_{k}\frac{C_{k}^{2}}{m_{k}\omega_{k}}\delta\left(\omega-\omega_{k}\right),\]
 that completely characterizes it. In our proposed model we have an
Ohmic case, eq {[}7{]}, given by \begin{equation}
\mathcal{J}\left(\omega\right)=\left\{ \begin{array}{c}
\eta\omega\,\, for\,\omega<\Omega\\
0\,\,\,\, for\,\omega>\Omega\end{array}\right.\end{equation}
 where we have introduced a cutoff frequency $\Omega$. With this
we obtain\begin{equation}
J_{ss'}\left(x,y,t;x',y',0\right)=\int_{x'}^{x}Dx\left(t'\right)\int_{y'}^{y}Dy\left(t'\right)\, e^{\frac{1}{\hbar}\beta_{ss'}\left[x\left(t'\right),y\left(t'\right)\right]}\end{equation}
with \begin{widetext}

\begin{eqnarray}
\beta_{ss'}\left[x\left(t'\right),y\left(t'\right)\right] & = & i\int_{0}^{t}dt'\,
\left[\frac{m}{2}\left(\dot{x}^{2}-\dot{y}^{2}\right)-f_{0}\left(t'\right)\left(Sx-S'y\right)-
\frac{\eta}{2}\left(Sx-S'y\right)\left(S\dot{x}+S'\dot{y}\right)\right]+\nonumber \\
 & - & \int_{0}^{t}dt'\int_{0}^{t'}dt''\,\left[Sx\left(t'\right)-S'y\left(t'\right)
 \right]\alpha_{R}\left(t'-t''\right)\left[Sx\left(t''\right)-S'y\left(t''\right)\right].
 \label{eq:}\end{eqnarray}

Computing the diagonal \(\left(S=S'=\pm1\right)\) and the
off-diagonals \(\left(S\neq S'=\pm1\right)\) terms we
obtain\begin{eqnarray}
\beta_{d}\left[x\left(t'\right),y\left(t'\right)\right] & = & i\int_{0}^{t}dt'\,\left[\frac{m}{2}\left(\dot{x}^{2}-\dot{y}^{2}\right)\mp f_{0}\left(t'\right)\left(x-y\right)-\frac{\eta}{2}\left(x-y\right)\left(\dot{x}+\dot{y}\right)\right]+\nonumber \\
 & - & \int_{0}^{t}dt'\int_{0}^{t'}dt''\,\left[x\left(t'\right)-y\left(t'\right)\right]\alpha_{R}\left(t'-t''\right)\left[x\left(t''\right)-y\left(t''\right)\right]\label{eq:}\end{eqnarray}
and

\begin{eqnarray}
\beta_{od}\left[x\left(t'\right),y\left(t'\right)\right] & = & i\int_{0}^{t}dt'\,\left[\frac{m}{2}\left(\dot{x}^{2}-\dot{y}^{2}\right)\mp f_{0}\left(t'\right)\left(x+y\right)-\frac{\eta}{2}\left(x+y\right)\left(\dot{x}-\dot{y}\right)\right]+\nonumber \\
 & - & \int_{0}^{t}dt'\int_{0}^{t'}dt''\,\left[x\left(t'\right)+y\left(t'\right)\right]\alpha_{R}\left(t'-t''\right)\left[x\left(t''\right)+y\left(t''\right)\right].\label{eq:}\end{eqnarray}

\end{widetext}So we have a different dynamics for the diagonal and off-diagonal
terms. Defining new coordinates $q=\left(x+y\right)/2$ and $\xi=x-y$
we decouple the $x\left(t\right)$ and $y\left(t\right)$ trajectories
and get\begin{eqnarray}
J_{d\left(od\right)}\left(q,\xi,t;q',\xi',0\right) & = & \int_{q'}^{q}Dq\left(t'\right)\int_{\xi'}^{\xi}D\xi\left(t'\right)\label{eq:25}\\
 &  & e^{\frac{1}{\hbar}\left[i\beta'_{d\left(od\right)}-\beta''_{d\left(od\right)}\right]}\nonumber \end{eqnarray}
 with

\begin{eqnarray}
\beta'_{d} & = & \int_{0}^{t}dt'\,\left[m\dot{q}\dot{\xi}-\eta\xi\dot{q}\mp f_{0}\left(t'\right)\xi\right],\\
\beta''_{d} & = & \int_{0}^{t}dt'\int_{0}^{t'}dt''\,\xi\left(t'\right)\alpha_{R}\left(t'-t''\right)\xi\left(t''\right),\end{eqnarray}
\begin{eqnarray}
\beta'_{od} & = & \int_{0}^{t}dt'\,\left[m\dot{q}\dot{\xi}-\eta\xi q\mp2f_{0}\left(t'\right)q\right]\end{eqnarray}

and

\begin{equation}
\beta''_{od}=4\int_{0}^{t}dt'\int_{0}^{t'}dt''\, q\left(t'\right)\alpha_{R}\left(t'-t''\right)q\left(t''\right).\end{equation}

Evaluating the integral in the usual way we expand it about the classical
trajectory and solve the remaining functional integral to obtain

\begin{equation}
J_{d\left(od\right)}\left(q,\xi,t;q',\xi',0\right)=e^{\frac{1}{\hbar}\left[i\beta_{d\left(od\right)}^{\prime\, class}-\beta_{d\left(od\right)}^{\prime\prime\, class}\right]}G\left(q,\xi,t;q',\xi',0\right)\end{equation}
 with

\begin{equation}
G\left(t\right)=\frac{m\gamma e^{\gamma t}}{2\pi\hbar\sinh\gamma t},\end{equation}

\begin{eqnarray}
\beta_{d}^{\prime\, class} & = & \xi qL_{-}\left(t\right)+\xi'q'L_{+}\left(t\right)-\xi q'N\left(t\right)\nonumber \\
 &  & -\xi'qM\left(t\right)\mp X\left(t\right)\xi\mp Z\left(t\right)\xi'\,,\label{eq:}\end{eqnarray}
\begin{eqnarray}
\beta_{od}^{\prime\, class} & = & \xi qL_{-}\left(t\right)+\xi'q'L_{+}\left(t\right)-\xi q'M\left(t\right)\nonumber \\
 &  & -\xi'qN\left(t\right)\mp2X\left(t\right)q\mp2Z\left(t\right)q'\,,\label{eq:}\end{eqnarray}

\begin{equation}
\beta_{d}^{\prime\prime\, class}=\frac{1}{2}\left[\xi^{2}A\left(t\right)+2\xi\xi'B\left(t\right)+\xi'^{2}C\left(t\right)\right],\end{equation}

and

\begin{equation}
\beta_{od}^{\prime\prime\, class}=2\left[q^{2}A\left(t\right)+2qq'B\left(t\right)+q'^{2}C\left(t\right)\right].\end{equation}

In the above expressions we have used $\xi\left(0\right)=\xi',\,\xi\left(t\right)=\xi$
and the same notation holds for $q$. We have also defined the following
functions\begin{eqnarray}
A\left(t\right) & = & \frac{e^{-2\gamma t}}{\sinh^{2}\gamma t}\int_{0}^{t}dt'\int_{0}^{t}dt''\, e^{\gamma\left(t'+t''\right)}\sinh\gamma t'\nonumber \\
 &  & \times\alpha_{R}\left(t'-t''\right)\sinh\gamma t'',\label{eq:}\end{eqnarray}
\\
\begin{eqnarray}
C\left(t\right) & = & \frac{1}{\sinh^{2}\gamma t}\int_{0}^{t}dt'\int_{0}^{t}dt''\, e^{\gamma\left(t'+t''\right)}\sinh\gamma\left(t-t'\right)\nonumber \\
 &  & \times\alpha_{R}\left(t'-t''\right)\sinh\gamma\left(t-t''\right),\label{eq:}\end{eqnarray}
\\
\begin{eqnarray}
B\left(t\right) & = & \frac{e^{-\gamma t}}{\sinh^{2}\gamma t}\int_{0}^{t}dt'\int_{0}^{t}dt''\, e^{\gamma\left(t'+t''\right)}\sinh\gamma t'\,\nonumber \\
 &  & \times\alpha_{R}\left(t'-t''\right)\sinh\gamma\left(t-t''\right),\label{eq:}\end{eqnarray}

\begin{equation}
L_{\pm}\left(t\right)=m\gamma\left[\coth\left(\gamma t\right)\pm1\right],\end{equation}
\\
\begin{equation}
N\left(t\right)=m\gamma\frac{e^{-\gamma t}}{\sinh\gamma t}\,,\end{equation}
\\
\begin{equation}
M\left(t\right)=m\gamma\frac{e^{\gamma t}}{\sinh\gamma t}\,,\end{equation}
\\
\begin{equation}
X\left(t\right)=\frac{e^{-\gamma t}}{\sinh\left(\gamma t\right)}\int_{0}^{t}dt'\, f_{0}\left(t'\right)e^{\gamma t'}\sinh\gamma t'\,,\end{equation}

and

\begin{equation}
Z\left(t\right)=\frac{1}{\sinh\left(\gamma t\right)}\int_{0}^{t}dt'\, f_{0}\left(t'\right)e^{\gamma t'}\sinh\gamma\left(t-t'\right).\end{equation}

For an initial Gaussian packet centered at the origin,\begin{equation}
\psi\left(x\right)=\frac{1}{\sqrt{\sqrt{2\pi}\sigma}}\,\, e^{-\frac{x^{2}}{4\sigma^{2}}}\,,\end{equation}
 we have

\begin{equation}
\rho\left(q',\xi',0\right)=\frac{1}{\sqrt{2\pi\sigma^{2}}}\,\, e^{-\frac{q'^{2}}{2\sigma^{2}}}\,\, e^{-\frac{\xi'^{2}}{8\sigma^{2}}}\,.\label{paocte inicial}\end{equation}

The diagonal element of the reduced density operator for the system
becomes\begin{widetext}\begin{eqnarray}
\rho_{d}\left(q,\xi,t\right) & = & \sqrt{\frac{\pi}{a}}G\left(t\right)exp\left\{ -\frac{M^{2}}
{4a\hbar^{2}}\left(q\mp\frac{Z}{M}\right)^{2}-\xi^{2}\left[\frac{1}
{2}\frac{A}{\hbar}+\frac{N^{2}\sigma^{2}}{2\hbar^{2}}-\frac{1}{4a\hbar^{4}}
\left(\sigma^{2}NL_{+}-\hbar B\right)^{2}\right]+\right.\nonumber \\
 &  & \left.+\frac{i}{\hbar}\left[L_{-}-\frac{M}{2a\hbar^{2}}
 \left(\sigma^{2}NL_{+}-\hbar B\right)\right]q\xi\pm\frac{i}{\hbar}
 \left[X+\frac{Z}{2a\hbar^{2}}\left(\sigma^{2}NL_{+}-\hbar B\right)\right]\xi\right\}
 \label{eq:47}\end{eqnarray}
with

\begin{equation}
a=\frac{1}{\hbar^{2}}\left(\frac{L_{+}^{2}\sigma^{2}}{2}+\frac{\hbar
C}{2}+\frac{\hbar^{2}}{8\sigma^{2}}\right),\end{equation}
whereas the off-diagonal element is

\begin{eqnarray}
\rho_{od}\left(q,\xi,t\right) & = & G\left(t\right)\sqrt{\frac{\pi}{a}}exp\left
\{ \left[-\frac{2\sigma^{2}N^{2}}{\hbar^{2}}\left(1-\frac{\sigma²L_{+}^{2}}{2a\hbar^{2}}\right)
+\frac{2}{\hbar}\left(\frac{B^{2}}{2a\hbar}-A\right)\right]q^{2}+\right.
\nonumber \\
 & + & \frac{i}{\hbar}\left[\left(\xi M\mp2Z\right)
 \left[\frac{B}{2a\hbar}-\frac{\sigma²NL_{+}}{2a\hbar^{2}}\right]+
 \left(\xi L_{-}\pm2X\right)\right]q+\nonumber \\
 & - & \left.\frac{\left(\xi M\mp2Z\right)^{2}}{16a\hbar^{2}}-\frac{2\sigma^{2}
 L_{+}NB}{a\hbar^{3}}q^{2}\right\}. \label{eq:}\end{eqnarray}

\end{widetext}So we have obtained an analytical expression for the reduced density
operator of the system in the spin space, when the initial spatial
part of the wave function is a Gaussian centered at the origin. Now
we are going to analyze this result.

\section{Spin coherence in the SGI}

The positions of the center of the packets are given by the probability
density\begin{equation}
\rho_{d}\left(q,0,t\right)=\frac{1}{\sqrt{2\pi}\tilde{\sigma}\left(t\right)}exp-\frac{1}{2\tilde{\sigma}\left(t\right)^{2}}\left[q\mp\frac{Z}{M}\right]^{2},\end{equation}
 with

\begin{equation}
\tilde{\sigma}\left(t\right)=\frac{\hbar\sqrt{2a\left(t\right)}}{M\left(t\right)}.\end{equation}

So, as expected, we have a Gaussian packet with width $\tilde{\sigma}\left(t\right)$
whose center is described by

\begin{equation}
z\left(t\right)=\frac{Z\left(t\right)}{M\left(t\right)}=\frac{1}{2m\gamma}\int_{0}^{t}dt'\, f_{0}\left(t'\right)\left(1-e^{-2\gamma\left(t-t'\right)}\right),\end{equation}
 which is the classical trajectory of a particle subject to a force
$f_{0}\left(t\right)$ and to the viscous force $\eta v$.

To see how pure the state is as it evolves in the SGI we will investigate
the off-diagonal elements of the density operator in the spin space.
In the high temperature limit, $KT\gg\hbar\gamma$, the functions
$A\left(t\right)$, $B\left(t\right)$ and $C\left(t\right)$ can
be easily evaluated and yield\begin{eqnarray}
\rho_{od}\left(t\right) & = & h\left(t\right)\exp\left\{ -\frac{1}{a'\hbar^{2}}\left[Z\left(\frac{B}{2a\hbar}-\frac{\sigma²NL_{+}}{2a\hbar^{2}}\right)-X\right]^{2}\right.\nonumber \\
 &  & \left.-\frac{1}{2}\left(\frac{\Delta\tilde{z}}{\tilde{\sigma}\left(t\right)}\right)^{2}\right\} \label{eq:53}\end{eqnarray}
 with

\begin{equation}
h\left(t\right)=\frac{M\left(t\right)}{2\hbar\sqrt{a\left(t\right)a'\left(t\right)}}\,\,\,\,\,\,\,\textrm{and}\end{equation}
\begin{eqnarray}
a'\left(t\right) & = & \frac{2\sigma^{²}N^{2}}{\hbar²}\left(1-\frac{\sigma²L_{+}^{2}}{2a\hbar^{2}}\right)-\frac{2}{\hbar}\left(\frac{B^{2}}{2a\hbar}-A\right)\nonumber \\
 &  & +\frac{2\sigma²L_{+}NB}{a\hbar^{3}}.\label{eq:55}\end{eqnarray}

In the non-dissipative limit, $\gamma\rightarrow0$, $h\left(t\right)$
goes to one and the exponential recovers the non-dissipative expression
for the reduced density operator. Therefore, we can say that the irreversible
loss of coherence is in the factor $h\left(t\right)$. In figures
3 and 4 we show the qualitative behavior of the exponential term and
the $h\left(t\right)$ factor.

\begin{figure}
\selectlanguage{brazil}
\includegraphics[scale=0.8]{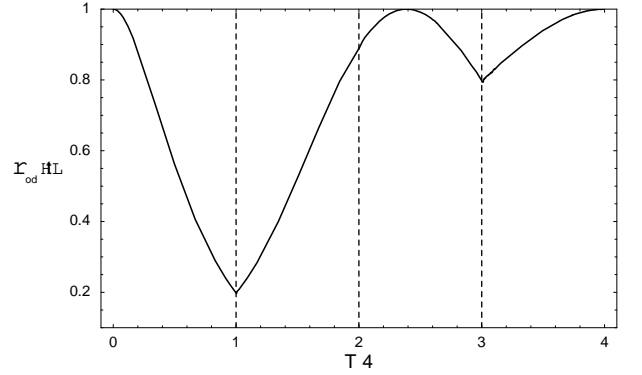}

\selectlanguage{english}

\caption{Behaviour of the exponent of (53), showing the possible recovery
of coherence. T
, the experiment time, is here of the order of $10^{-9}s$. }
\end{figure}

\begin{figure}
\selectlanguage{brazil}
\includegraphics[scale=0.8]{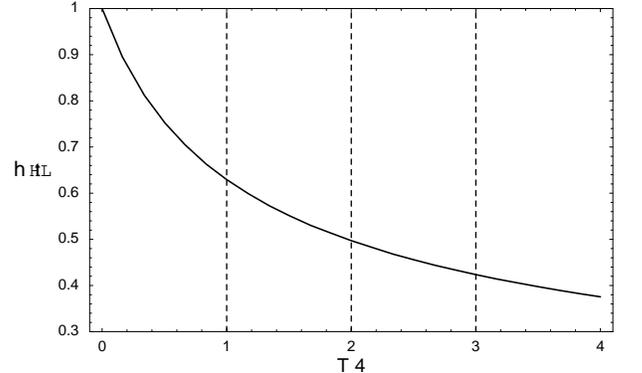}

\selectlanguage{english}

\caption{Factor $h\left(t\right)$ showing the irreversible loss of coherence.
T is the same as above.}
\end{figure}

In these graphs we can see that the recombination could lead to a
revival of coherence, but $h\left(t\right)$ continually decreases
which makes this revival negligible as shown in figure 5, where, in
the end, we have only $20$\% of coherence left.

\begin{figure}
\selectlanguage{brazil}
\includegraphics[scale=0.8]{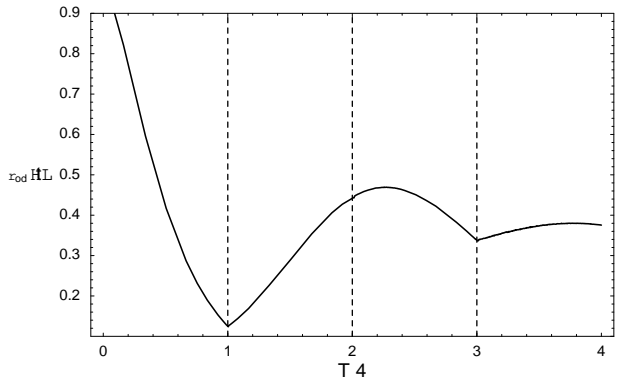}

\selectlanguage{english}

\caption{Behaviour of the non-diagonal element in the spin space.}
\end{figure}

Despite the qualitative nature of the graphs, we can say that in order
to have a significant revival of coherence the time of the experiment
should be much shorter than the decoherence time, which can be extracted
from the expression for $h\left(t\right)$. Since this expression
is very cumbersome, we have to do it grafically.

For the proposed model of a magnetic field generated by a pair of
SQUIDs with dimensions of microns, we obtained $\gamma^{-1}=10^{15}s$.
These values and the other parameters already used give us the decoherence
time, $\tau$, of $10^{5}s$ for temperatures close to $0.1K$. 

In the usual Stern-Gerlach experiment the particles have velocities
of about $1000\, m/s$ that give us an interaction time of the order
of $10^{-6}s$, which is smaller than the decoherence time. So, for
particles with velocities of the order of $1000\, m/s$ decoherence
can not be observed and will not be important in the SGI. Therefore,
in order to control the spin coherence time in an observable way we
should use SQUIDs of smaller dimensions which will furnish us with
shorter relaxations times. Larger SQUIDs would only make matters even
worse.

Although we have only stressed the possible effect of field fluctuations
on the recombination one should bear in mind that the perfect recombination
is by itself a very difficult task to perform.

\section{Summary}

In this work we have investigated the effect of spin coherence, discussing
experiments with beams of particles subject to magnetic fields. We
have extended the analysis of the Stern-Gerlach interferometer adding
unavoidable fluctuations of the magnetic field through the coupling
of the system to a bath of harmonic oscillators. The introduction
of these fluctuations adds a factor to the off-diagonal elements of
the reduced density operator in the spin space that decays in a time
$\tau$, the decoherence time. However we have shown that even for
a very {}``noisy'' environment provided by particularly chosen parameters
for a SQUID the decoherence time is still extremely long compared
to the experiment time.

When the magnetic field is generated by our proposed micrometric SQUID,
we observed that dissipative effects are not important when the particles
velocities are of the order of $1000\, m/s$. We believe that changing
the SQUID parameters it will be possible to observe the loss of spin
coherence by reducing the SQUID size. In this case we will have a
model of a reservoir for a genuine microscopic quantum mechanical
variable which, in our case, is the magnetic moment of an atom!

\begin{acknowledgments}
T. R. O., is grateful for FAPESP (Fundação de Amparo à Pesquisa do
Estado de São Paulo) whereas A.O.C. acknowledges the support from
CNPq (Conselho Nacional de Desenvolvimento Científico e Tecnológico)
and the Millennium Institute for Quantum Information.
\end{acknowledgments}

\end{document}